\documentclass{ws-ijqi_pdf}

\begin{document}


%
%

\title{FLUCTUATIONS OF QUANTUM RANDOM WALKS ON CIRCLES}

\author{NORIO INUI, YOSHINAO KONISHI$^{\dagger}$}

\address{Department of Mechanical and Intelligent Engineering, Himeji Institute of Technology, \\
2167 Syosha, Himeji, 671-2203, Japan\\
inui@mie.eng.himeji-tech.ac.jp, tm02m018@mie.eng.himeji-tech.ac.jp$^{\dagger}$}

\author{NORIO KONNO, TAKAHIRO SOSHI$^{\ast}$}

\address{Department of Applied Mathematics, 
Yokohama National University, \\
79-5 Tokiwadai, Yokohama, 240-8501, Japan\\
norio@mathlab.sci.ynu.ac.jp, soshi@lam.osu.sci.ynu.ac.jp$^{\ast}$}

\maketitle

\begin{history}
\end{history}

\begin{abstract}
Temporal fluctuations in the Hadamard walk on circles are studied.
A temporal standard deviation of probability that a quantum random 
walker is positive at a given site is introduced to manifest striking 
differences between quantum and classical random walks. 
An analytical expression of the temporal standard deviation on a circle 
with odd sites is shown and its asymptotic behavior is considered 
for large system size. In contrast with classical random walks, 
the temporal fluctuation of quantum random walks depends 
on the position and initial conditions, 
since temporal standard deviation of the classical case is zero for any site. 
It indicates that the temporal fluctuation of the Hadamard walk can be controlled.
\end{abstract}


\section{Introduction}
\hspace*{1em}
Classical random walks have been used as an important tool in modern computers. It is realized by accumulating results obtained from mathematics and computer science. Quantum computer has not appeared in satisfactory form, however it is valuable to study both discrete and continuous quantum random walks to seek novel applications of quantum computers for the future. We refer to Ref. 1 and the references therein, concerning quantum random walks. 

In this paper we focus on discrete time case. Studies of quantum random walks revealed common properties and differences between classical and quantum random walks. If a classical random walker start from the origin on an infinite line, the spatial standard deviation of distribution increases in proportion to $\sqrt{t}$ after $t$ steps. On the other hand, a spatial standard deviation of distribution in the Hadamard walk whose initial state has a value only at the origin increases in proportion to $t$ after sufficiently large $t$ steps. See Refs. 2 and 3, for examples.

From now on we consider quantum random walks on circles. The probability distribution of a classical random walker on a circle with odd sites becomes uniform in the limit of $t \rightarrow \infty$, moreover time average of the probability distribution on a circle with both even and odd sites converges the uniform distribution. However, the probability distribution of quantum random walk does not converge to any stationary state even for odd case. On the other hand, its time average converges to uniform distribution for odd sites and does not converge to uniform one for even sites.$^{4,5}$

For that reason time-averaged distribution of quantum random walk, which converges a constant at a fixed site was introduced. It should be kept in our mind that the probability of being at position $n$ fluctuates always bordering on time-averaged density. In the case of the Hadmard walk, the time-averaged distribution  becomes uniform independently on the initial state if the number of site is odd, and it is agreed with the classical random walk. The fluctuation of distribution is not mere mathematical object, however it is related to measurement.
The time-averaged distribution is obtained by measuring the state at a random time chosen in a certain interval. If the fluctuation of density function is small, we can expect to obtained good time-averaged distribution by the few number of measurements. 

Although details of the time-averaged distribution of the Hadamard walk on a circle are known, (see Refs. 4 and 5, for examples), little about the fluctuation of the quantum random walks is known. In this paper we introduce a temporal standard deviation to characterize the fluctuation of quantum random walk and show that the temporal standard deviation corresponding to the Hadamrd walk depends on the location of sites even if time-averaged density function is uniform. This result has a striking difference between quantum and classical cases, since temporal standard deviation of the classical case is zero for any site. Furthermore we consider the dependence of the standard deviation on the initial state.

This paper is organized as follows. Sec. 2 gives the definition of the Hadamard walk and some preliminary results. In Sec. 3, we introduce a temporal standard deviation $\sigma_N (n)$ and compute it explicitly. Sec. 4 is devoted to a dependence of $\sigma_N (n)$ on the position and system size. Furthermore, Sec. 5 contains a dependence of $\sigma_N (n)$ on initial state and parity of system size. Finally in Sec. 6 we summarize our results.

\section{Hadamard Walk}
\hspace*{1em}
Let us consider a quantum random walker who moves on cyclic sites following the Hadamard transform. Suppose that there are $N$ sites on a circle and each site is labeled by an integer from 0 and $N-1$. In the Hadamard walk, the quantum state on a fixed site is characterized by ``chirality" descried ``left" and ``right".  In this paper, the site labeled by $n-1$ mod $N$ is regarded as   the site in the left to the $n$-th site.
To the contrary, the site labeled by  $n+1$ mod $N$ is regarded as  the site in the right to the $n$-th site.
Let $|L,n,t \rangle$ and $|R,n,t \rangle$ be the wave function on $n$-th site 
at time $t$ corresponding ``left" and ``right", respectively.
The time evolution of these wave functions is given by the Hadamard transform, that is, the transformation of the wave function for one step is descried by
\begin{eqnarray}
|L,n,t \rangle &=&\frac{1}{\sqrt{2}} |L,n+1,t-1 \rangle+\frac{1}{\sqrt{2}} |R,n+1,t-1 \rangle, \\
|R,n,t \rangle &=&\frac{1}{\sqrt{2}} |L,n-1,t-1 \rangle-\frac{1}{\sqrt{2}} |R,n-1,t-1 \rangle.
\label{Htrans}
\end{eqnarray}
Then the total state $\Psi(t) \equiv (|L,0,t \rangle,|R,0,t \rangle,
|L,1,t \rangle,|R,1,t \rangle,\cdots ,|L,N-1,t \rangle,|R,N-1,t \rangle )^{T}$
is transformed by the following unitary matrix with 2$N$ $\times$ 2$N$ elements,
\begin{eqnarray}
M_{N}=\left[ 
\begin{array}{cccccccc}
0 & P & 0 & \cdots & \cdots & 0 & Q \\
Q & 0 & P & 0 & \cdots & \cdots & 0 \\
0 & Q & 0 & P & 0 & \cdots & 0 \\
\vdots & \ddots & \ddots & \ddots & \ddots & \ddots & \vdots \\
0 & \cdots & 0 & Q & 0 & P & 0\\
0 & \cdots & \cdots & 0 & Q & 0 & P\\
P & 0 & \cdots & \cdots & 0 & Q & 0\\
\end{array} 
\right],
\end{eqnarray} 
where $T$ denotes transposition and 
\begin{equation}
P=\frac{1}{{\sqrt{2}}}\left[ 
\begin{array}{cc}
1 & 1 \\
0 & 0 \\
\end{array} 
\right], \quad
Q=\frac{1}{{\sqrt{2}}}\left[ 
\begin{array}{cc}
0 & 0 \\
1 & -1 \\
\end{array}
\right], \quad
0=\left[
\begin{array}{cc}
0 & 0 \\
0 & 0 \\
\end{array}
\right].
\end{equation} 
Thus the total state after $t$ step, $\Psi(t)$, is given by $M_{N}^{t}\Psi(0)$
for the initial state $\Psi(0)$. 

To express the total state as a function of $n$ and $t$, 
 we use known results on eigenvalues (Lemma 1) and eigenvectors (Lemma 2) of $M_{N}$ obtained by Aharonov et al.$^{4}$ and Bednarska et al.$^{5}$ 

\begin{lemma}
For $j = 0, 1, \ldots, N-1$ and $k=0,1$, the $(2j+k+1)$-th eigenvalue corresponding to $M_{N}$ is given by
\begin{eqnarray}
c_{jk} &=& {1 \over \sqrt{2}} \left\{ (-1)^k \, 
\sqrt{1 + \cos ^2 \left( \frac{2\,j\,\pi }{N} \right) } 
+i \sin \left( \frac{2\,j\,\pi }{N} \right) \right\}. 
\label{eqn:eigenV}
\end{eqnarray}
\end{lemma}

We here define $v_{j,k,l}^{o}$ and $v_{j,k,l}^{e}$ to express eigenvectors by
\begin{eqnarray}
v_{j,k,l}^{o} &=&{a_{jk}}\,{b_{jk}}\,{{{\omega}^{jl}_N}}, \\
v_{j,k,l}^{e} &=&{a_{jk}}\,{{{\omega}^{jl}_N}},
\label{veo}
\end{eqnarray}
where
\begin{eqnarray}
\omega_{N} &=& e^{\frac{2\,i\,\pi }{N}}, \\
a_{jk} &=&\frac{1}{{\sqrt{N \left( 1 + \left|{b_{jk}}\right| ^2 \right)}}},\\
b_{jk} &=&{{{{\omega}^{j}_N}}} 
\left\{ (-1)^k \,
 \sqrt{1 + \cos^2 \xi_j} +\cos \xi_j \right\}, \\
\xi_j &=& {2 j \pi \over N}.
\end{eqnarray}

\begin{lemma}
For $l=1,2, \ldots ,2N$, the $l$-th element of the eigenvectors corresponding to $c_{jk}$ is given by
\begin{eqnarray}
v_{j,k,l}=\left\{ 
\begin{array}{ll}
v_{j,k,(l+1)/2}^{o} & (l=odd), \\
\\
v_{j,k,l/2}^{e} & (l=even), \\
\end{array} \right.
\label{eqn:eigenVec}
\end{eqnarray}
\end{lemma}

For the convenience of readers, from now on here we check directly that eigenvectors $v_{j,k}=(v_{j,k,1},\cdots,v_{j,k,2N})^{T}$ satisfy the next equation
\begin{equation}
M_{N} v_{j,k}=c_{j,k} v_{j,k}.
\label{eqn:defeigen}
\end{equation}


(a) Suppose $l$ is odd, then the $(2m-1)$-th element  
of the left-hand side of the Eq. (\ref{eqn:defeigen}) for $m=1,2,\cdots,N$ is given by
\begin{eqnarray}
&&(2m-1) \mbox{-th element of LHS}  \nonumber\\
&=&\frac{1}{\sqrt{2}} 
\left(v_{j,k,m+1 \mbox{\,mod\,} N}^{o}+v_{j,k,m+1 \mbox{\,mod\,} N}^{e} \right), \nonumber\\
&=&      \frac{ a_{jk}}{\sqrt{2}}
          \left[ 
           e^{\frac{2j(m+1)i\pi}{N}} 
          \left\{
          1+ e^{\frac{2ji\pi}{N}}
          \left((-1)^k
          \sqrt{1+\cos^{2} \xi_j} 
          + \cos \xi_j
          \right)
          \right\}
          \right],
\label{eqn:L1}
\end{eqnarray}
where $\xi_j=2 \pi j/N$. On the other hand, the right-hand side of Eq. (\ref{eqn:defeigen}) is given by
\begin{eqnarray}
&&(2m-1) \mbox{-th element of RHS}  \nonumber\\
&=& c_{jk}a_{jk}b_{jk}\omega^{jm}_N, \nonumber \\
        &=& \frac{ a_{jk}}{\sqrt{2}}
          \left[ 
           e^{\frac{2j(m+1)i\pi}{N}} 
          \left( (-1)^k
          \sqrt{1+\cos^{2} \xi_j} 
          + \cos \xi_j
          \right)
          \right.  \nonumber \\ 
&&        \left.
          \hspace{23mm} \times
          \left(  (-1)^k
          \sqrt{1+\cos^{2} \xi_j}+i \sin \xi_j
          \right)
          \right].
\label{eqn:L2}
\end{eqnarray}

(b) Similarly suppose $l$ is even, then the $2m$-th element of the left-hand side 
of the Eq. (\ref{eqn:defeigen})  for $m=1,2,\cdots,N$ is given by
\begin{eqnarray}
&&2m \mbox{-th element of LHS}  \nonumber\\
&=&\frac{1}{\sqrt{2}} \left(v_{j,k,m-1 \mbox{\,mod\,} N}^{o}-v_{j,k,m-1 \mbox{\,mod\,} N}^{e} \right),  \nonumber \\
      &=& \frac{ a_{jk}}{\sqrt{2}}
          \left[ 
           e^{ \frac{2j(m-1)i\pi}{N} } 
          \left\{  
          -1+ e^{ \frac{2ji\pi}{N} }
          \left( (-1)^k 
          \sqrt{1 + \cos^{2} \xi_j} 
          + \cos \xi_j 
          \right)
          \right\}
          \right].
\label{eqn:R1}
\end{eqnarray}

On the other hand, the right-hand side of the Eq. (\ref{eqn:defeigen}) is given by
\begin{eqnarray}
&&2m \mbox{-th element of RHS}  \nonumber\\
&=& c_{jk}a_{jk}\omega^{jm} _N, \nonumber \\
      &=& \frac{ a_{jk}}{\sqrt{2}}
          \left[ 
           e^{\frac{2jmi\pi}{N}} 
          \left(
          (-1)^k \sqrt{1+\cos^{2} \xi_j}+i \sin \xi_j
          \right)
          \right].
\label{eqn:R2}
\end{eqnarray}
Using the Eular's formula, we can check that both sides of Eq.(\ref{eqn:defeigen}) coincide for each case.  

\section{Temporal Standard Deviation}
\hspace*{1em}
Since we obtained the eigenvalues and eigenvectors, matrix $M_{N}$ is transformed into a diagonal matrix and the wave function after $t$ steps is generally expressed by a liner 
combination
of $c_{jk}^{t}$. In real system we can not observe the wave function but probability of being at position $n$ at time $t$. This probability $P_{N}(n,t)$ is calculated from  the wave functions by
\begin{eqnarray}
P_{N}(n,t)=\langle L,n,t|L,n,t \rangle+\langle R,n,t |R,n,t \rangle.
\label{eqn:P}
\end{eqnarray}

In contrast to classical random walks, 
the probability $P_{N}(n,t)$ dose not converge in the limit of $t \rightarrow \infty$.
Thus we consider a time-averaged distribution defined by
\begin{eqnarray}
\bar{P}_{N}(n)=\lim_{T \rightarrow \infty} \frac{1}{T} \sum_{t=0}^{T-1} P_{N}(n,t),
\label{eqn:average}
\end{eqnarray}
if the right-hand side of Eq. (\ref{eqn:average}) exists. Aharonov et al. $^{4}$ showed that $\bar{P}_{N}(n)$ exists, and it is independent of both an 
initial state and $n$ if all eigenvalues of $M_{N}$ are distinct.
If the number of site is odd, then all eigenvalues of $M_{N}$ are distinct, so we immediately have $\bar{P}_{N}(n)=1/N$ for any $n=0,1, \ldots, N-1$. On the other hand, if $N$ is even, then there exist degenerate eigenvalues of 
$M_{N}$ and it is possible to derive non-uniform distributions.
In the case of the classical random walk, the time-averaged 
distribution of a walker becomes uniform
independently on the parity of the system size. Thus we can not distinguish the classical random walk and the Hadamard walk on a circle with odd sites only by the time-averaged distribution. 

As we mentioned above, the $P_{N}(n,t)$ does not converge
in the limiting $t \rightarrow \infty$.
It means that the probability always fluctuates to the upper and lower sides of $P_{N}(n)$.
For this reason, we define the following temporal standard deviation $\sigma_{N}(n)$:
\begin{eqnarray}
\sigma_{N}(n)=\sqrt{
\lim_{T \rightarrow \infty} \frac{1}{T} 
\sum_{t=0}^{T-1} \left (P_{N}(n,t)-\bar{P}_{N}(n) \right )^{2}
},
\label{eqn:sigma}
\end{eqnarray}
if the right-hand side of Eq. (\ref{eqn:sigma}) exists. 

Here we consider the classical case. In the case of classical a random walk starting from a site for odd sites (i.e., aperiodic case), there exist $a \in (0,1)$ and $C >0$ (are independent of $n$ and $t$) such that
\begin{eqnarray*}
 |P_{N}(n,t)-\bar{P}_{N}(n)| \le C a^t,
\end{eqnarray*}
where $\bar{P}_{N}(n)=1/N$ for any $n=0,1, \ldots, N-1$, (see page 63 of Schinazi,$^{6}$ for example). Therefore we obtain
\begin{eqnarray*}
\frac{1}{T} \sum_{t=0}^{T-1} \left (P_{N}(n,t)-\bar{P}_{N}(n) \right )^{2}
\le {C^2 \over T} {1-a^{2T} \over 1-a^2}.
\end{eqnarray*}
The above inequality implies that for any $n=0,1, \ldots, N-1$,
\begin{eqnarray*}
\sigma_{N}(n)=0.
\end{eqnarray*}
in the classical case. As for $N=$ even (i.e., periodic) case, we have the same conclusion $\sigma_{N}(n)=0$ for any $n$ by using a little modified argument.

In this situation, we first ask a natural question whether the $\sigma_{N}(n)$ is  always zero or not in quantum case. Moreover if $\sigma_{N}(n)$ is not always zero, then we next ask a question whether the $\sigma_{N}(n)$ is uniform or not.

From now on we concentrate our attention to the system with odd sites. Furthermore we assume that the initial state is a fixed $\Psi(0)=(1,0,0,\cdots ,0)^T$.
Let us express the standard deviation $\sigma_{N}(n)$ starting $\Psi(0)=(1,0,0,\cdots , 0)^T$ by
wave functions. To do so, plugging Eq. (\ref{eqn:P}) into Eq.(\ref{eqn:sigma}) we have
\begin{eqnarray}
\sigma_{N}^2 (n) = \lim_{T \rightarrow \infty} \frac{1}{T} 
\sum_{t=0}^{T-1} \left (\langle L,n,t|L,n,t \rangle+\langle R,n,t |R,n,t \rangle
-\frac{1}{N} \right )^{2},
\label{eqn:Tsigma1}
\end{eqnarray}
where we used the result $\bar{P}(n)=1/N$ for odd $N$.
Since the matrix $M_{N}$  was already diagonalzed in the previous section, 
the wave function $|L,n,t \rangle$ and $|R,n,t \rangle$ are easily obtained as
\begin{eqnarray}
|L,n,t \rangle &=&\sum_{j=0}^{N-1} 
\left(
\alpha_{j0}c_{j0}^{t}+\alpha_{j1}^{}c_{j1}^{t} 
\right), \\
|R,n,t \rangle &=&\sum_{j=0}^{N-1} 
\left(
\beta_{j0}c_{j0}^{t}+\beta_{j1}^{}c_{j1}^{t} 
\right).
\label{eqn:wF}
\end{eqnarray}
where
\begin{eqnarray}
\alpha_{jk}&=&
{ e^{\frac{2nj \pi i}{N}} \over 2N }
\left(
 1+ (-1)^k
 \frac{
       \cos(\frac{2j\pi}{N})
      }
      {
       \sqrt{1+\cos^{2}(\frac{2j\pi}{N})}
      }
\right), 
\label{eqn:ab1}
\\
\beta_{jk} &=&
 \frac{ (-1)^k 
       e^{\frac{2(n-1)j \pi i}{N}}
      }
      {
       2N\sqrt{1+\cos^{2}(\frac{2j\pi}{N})}
      }.
\label{eqn:ab4}
\end{eqnarray}
Since the matrix $M_{N}$ is unitary matrix, the eigenvalue $c_{jk}$ is written as
$e^{i \theta_{jk}}$ where $\theta_{jk}$ is  the argument of $c_{jk}$. Thus the probability
$\langle L,n,t|L,n,t \rangle$ is expressed by
\begin{eqnarray}
\langle L,n,t|L,n,t \rangle &=& \sum_{j=0}^{N-1} ( |\alpha_{j0}|^{2}+|\alpha_{j1}|^{2}) \nonumber \\
&&
+\sum_{j_{0},j_{1}=0}^{N-1}
 \sum_{k_{0},k_{1}=0}^{1}
\delta_{j_{0}j_{1},k_{0}k_{1}}
\alpha_{j_{0}k_{0}}\alpha_{j_{1}k_{1}}^{\ast}e^{i(\theta_{j_{0}k_{0}}-\theta_{j_{1}k_{1}})t},
\label{eqn:LL}
\end{eqnarray}
where 
\begin{eqnarray}
\delta_{j_{0}j_{1},k_{0}k_{1}}
=\left\{
\begin{array}{cc}
0  & \hspace{3mm} 
j_{0}=j_{1} \hspace{3mm} \mbox{and} \hspace{3mm} k_{0}=k_{1} \\
1  & \mbox{otherwise}.
\end{array}
\right.
\end{eqnarray}

The first term of Eq. (\ref{eqn:LL}) is a constant and the second term vanishes by applying an
operation $\lim_{T \rightarrow \infty} \frac{1}{T}\sum_{t=0}^{T-1}$. Similarly we can divide
$\langle R,n,t|R,n,t \rangle$ into a constant term and a vanishing term:
\begin{eqnarray}
\langle R,n,t|R,n,t \rangle &=& \sum_{j=0}^{N-1} ( |\beta_{j0}|^{2}+|\beta_{j1}|^{2}) \nonumber \\
&&
+\sum_{j_{0},j_{1}=0}^{N-1}
 \sum_{k_{0},k_{1}=0}^{1}
\delta_{j_{0}j_{1},k_{0}k_{1}}
\beta_{j_{0}k_{0}}\beta_{j_{1}k_{1}}^{\ast}e^{i(\theta_{j_{0}k_{0}}-\theta_{j_{1}k_{1}})t}.
\label{eqn:RR}
\end{eqnarray}
As a result we have
\begin{eqnarray}
\bar{P}(n)=\sum_{j=0}^{N-1} ( |\alpha_{j0}|^{2}+|\alpha_{j1}|^{2}+|\beta_{j0}|^{2}+|\beta_{j1}|^{2}).
\label{eqn:P2}
\end{eqnarray}
Plugging Eqs. (\ref{eqn:ab1})-(\ref{eqn:ab4}) into Eq. (\ref{eqn:P2}) gives confirm $\bar{P}(n)=1/N$ for any $n=0,1, \ldots , N-1$.

Let express $\sigma_{N}^{2}(n)$ as function of eigenvalues. 
By using Eqs. (\ref{eqn:Tsigma1}), (\ref{eqn:LL}) and (\ref{eqn:RR}), we get
\begin{eqnarray}
\sigma_{N}^{2}(n) &=&
\lim_{T \rightarrow \infty} \frac{1}{T} 
\sum_{t=0}^{T-1}
\sum_{j_{0},j_{1},j_{2},j_{3}=0}^{N-1}
 \sum_{k_{0},k_{1},k_{2},k_{3}=0}^{1}
\delta_{j_{0}j_{1},k_{0}k_{1}}\delta_{j_{2}j_{3},k_{2}k_{3}} 
\nonumber \\
&& 
\times
(
\alpha_{j_{0}k_{0}}\alpha_{j_{1}k_{1}}^{\ast}
+
\beta_{j_{0}k_{0}}\beta_{j_{1}k_{1}}^{\ast}
)  
(
\alpha_{j_{2}k_{2}}\alpha_{j_{3}k_{3}}^{\ast}
+
\beta_{j_{2}k_{2}}\beta_{j_{3}k_{3}}^{\ast}
)e^{i \Delta \theta t}
,
\label{eqn:sigmaA}
\end{eqnarray}
where
\begin{eqnarray}
\Delta \theta & \equiv& \Delta \theta(j_{0},k_{0},j_{1},k_{1},j_{2},k_{2},j_{3},k_{3}) \nonumber \\
       &=& \theta_{j_{0}k_{0}}-
\theta_{j_{1}k_{1}}+
\theta_{j_{2}k_{2}}-
\theta_{j_{3}k_{3}}.
\label{eqn:Delta}
\end{eqnarray}

If $\Delta \theta \not= 0$ (mod 2$\pi$), then  
$\lim_{T \rightarrow \infty} \sum_{t=0}^{T-1} e^{i \Delta \theta t}/T$ converges to a zero.
Thus the variance $\sigma_{N}^{2}(n)$ is obtained by taking summation of
$
(
\alpha_{j_{0}k_{0}}\alpha_{j_{1}k_{1}}^{\ast}
+
\beta_{j_{0}k_{0}}\beta_{j_{1}k_{1}}^{\ast}
)  
(
\alpha_{j_{2}k_{2}}\alpha_{j_{3}k_{3}}^{\ast}
+
\beta_{j_{2}k_{2}}\beta_{j_{3}k_{3}}^{\ast}
)
$
over combinations $j_{0}, k_{0}, \cdots,j_{3}, k_{3}$ satisfying $\Delta \theta=0$ (mod 2$\pi$).
For this reason, we consider the combinations which satisfy 
$\Delta \theta=0$ (mod 2$\pi$) 
for a given combination $(j_{0}, k_{0}, j_{1}, k_{1})$.
We define the following four conditions:
\begin{eqnarray}
&(a)& \hspace{3mm}
\mbox{Re}
[c_{j_{0}k_{0}}] =  
\mbox{Re}
[c_{j_{1}k_{1}}] 
\hspace{5.7mm}
\mbox{and}
\hspace{3mm}
\mbox{Im}
[c_{j_{0}k_{0}}] = 
-\mbox{Im}[c_{j_{1}k_{1}}],  
\nonumber
\\
&(b)& \hspace{3mm}
\mbox{Re}
[c_{j_{0}k_{0}}] =-  
\mbox{Re}
[c_{j_{1}k_{1}}] 
\hspace{3mm}
\mbox{and}
\hspace{3mm}
\mbox{Im}
[c_{j_{0}k_{0}}] =  
\mbox{Im}[c_{j_{1}k_{1}}],  
\nonumber
\\
&(c)& \hspace{3mm}
\mbox{Re}
[c_{j_{0}k_{0}}] =-  
\mbox{Re}
[c_{j_{1}k_{1}}] 
\hspace{3mm}
\mbox{and}
\hspace{3mm}
\mbox{Im}
[c_{j_{0}k_{0}}] = 
-\mbox{Im}[c_{j_{1}k_{1}}], 
\nonumber 
\\
&(d)& \hspace{3mm}
\mbox{Re}
[c_{j_{0}k_{0}}] =  
\mbox{Re}
[c_{j_{1}k_{1}}] 
\hspace{5.7mm}
\mbox{and}
\hspace{3mm}
\mbox{Im}
[c_{j_{0}k_{0}}] = 
\mbox{Im}[c_{j_{1}k_{1}}].  
\hspace{3mm} 
\label{eqn:condition1}
\end{eqnarray}
If each condition from (a) to (d) is not satisfied, there are four different combinations 
which satisfy $\Delta \theta=0$ (mod 2$\pi$) 
for a given combination $(j_{0},k_{0},j_{1},k_{1})$:
\begin{eqnarray}
&(a)& \hspace{3mm}
j_{2}=N-j_{0} \hspace{2mm} \mbox{mod} \hspace{2mm} N,
\hspace{2mm}
k_{2}=k_{0},
\hspace{2mm}
j_{3}=N-j_{1} \hspace{2mm} \mbox{mod} \hspace{2mm} N,
\hspace{2mm}
k_{3}=k_{1}, \nonumber
\\
&(b)& \hspace{3mm}
j_{2}=j_{0},
\hspace{2mm}
k_{2}=1-k_{0},
\hspace{2mm}
j_{3}=j_{1},
\hspace{2mm}
k_{3}=1-k_{1}, \nonumber
\\
&(c)& \hspace{3mm}
j_{2}=N-j_{1} \hspace{2mm} \mbox{mod} \hspace{2mm} N,
\hspace{2mm}
k_{2}=1-k_{1},
\hspace{2mm}
j_{3}=N-j_{0} \hspace{2mm} \mbox{mod} \hspace{2mm} N,
\hspace{2mm}
k_{3}=1-k_{0},  \nonumber
\\
&(d)& \hspace{3mm}
j_{2}=j_{1},
\hspace{2mm}
k_{2}=k_{1},
\hspace{2mm}
j_{3}=j_{0},
\hspace{2mm}
k_{3}=k_{0}.
\label{eqn:comb1}
\end{eqnarray}
If one of condition in (\ref{eqn:condition1}) is satisfied, we find the same values in (\ref{eqn:comb1}).

We can introduce a geometrical picture into these combinations.
All eigenvalues exist on a unit circle in a complex plane and
there is a symmetry in these points. 
For example, when  we find $c_{j_{0}k_{0}}$ at the point $A$,
we can find always another eigenvalue 
on a point which is obtained by reflecting  point $A$ in the $x-$axis and
its evienvalues is expressed by $c_{N-j_{0} \mbox{\,mod\,} N,k_{0}}$.
We also find always an eigenvalue 
on a point which is obtained by reflecting  point $A$ in the $y-$axis and
its eigenvalues is expressed by $c_{j_{0},1-k_{0}}$. Additionally, we note that
if the relation between $c_{j_{0}k_{0}}$  and $c_{j_{1}k_{1}}$ is  
symmetrical with respect to the origin and 
the relation between $c_{j_{2}k_{2}}$  and $c_{j_{3}k_{3}}$ is also  
symmetrical with respect to the origin, then $\Delta \theta = 0$ (mod 2$\pi$).

From the relation (\ref{eqn:comb1}) we express $j_{2}$,$k_{2}$,$j_{3}$ and $k_{3}$ as
a function of $j_{0}$, $k_{0}$, $j_{1}$ and $k_{1}$. Thus
$\sigma_{N}^{2}(n)$ is obtained by summations over $j_{0}$, $k_{0}$, $j_{1}$ and  $k_{1}$.
Taking the degenerate into account correctly we obtain $\sigma_{N}^{2}(n)$ by some computations. Before showing the result we define next functions 
\begin{eqnarray}
S_{0} &=& \sum_{j=0}^{N-1} \frac{1}{3+\cos \theta_{j} }, \\
S_{1} &=& \sum_{j=0}^{N-1} \frac{\cos \theta_{j}}{3+\cos \theta_{j}}, \\
S_{+}(n) &=& \sum_{j=0}^{N-1} \frac{  \cos \left(  (n-1) \theta_{j} \right)+\cos \left(n \theta_{j} \right) }
                  {3+\cos \theta_{j}}, \\
S_{-}(n) &=& \sum_{j=0}^{N-1} \frac{  \cos \left(  (n-1)\theta_{j} \right)-\cos \left( n \theta_{j}  \right)}
                  {3+\cos  \theta_{j} }, \\
S_{2}(n) &=& \sum_{j=1}^{N-1} \frac{7+\cos   2 \theta_{j} 
                                    +8\cos     \theta_{j}\cos^{2} 
                                            \left[ \left(n-\frac{1}{2} \right)
                                             \theta_{j}  \right] }
                                   {(3+\cos \theta_{j})^{2}},
\label{functions}
\end{eqnarray}
where $\theta_{j} \equiv 4 \pi j/N$.
Now we show  $\sigma_{N}^{2}(n)$ as a main result in this paper:

\begin{theorem} When $N$ is odd, we have
\begin{eqnarray}
\sigma_{N}^{2}(n)=\frac{1}{N^{4}} 
\left[
2 
\left\{ S_{+}^{2}(n)+S_{-}^{2}(n) \right\}
+11S_{0}^{2}+10 S_{0}S_{1}+3S_{1}^{2}-S_{2}(n)
\right]
-\frac{2}{N^{3}},
\label{eqn:sigmaF}
\end{eqnarray}
for any $n=0,1, \ldots, N-1$.
\end{theorem}

\section{Dependence of $\sigma_{N}(n)$ on the position and system size}
\hspace*{1em}
The obtained formula for $\sigma_{N}(n)$ is written in the form of a single summation, therefore,
it is calculated very easily in comparison with Eq. (\ref{eqn:sigmaA}) containing nine-fold summations.
We show  $\sigma_{N}(n)$ for $N=3,5,\cdots,11$ in Fig. 1.

\begin{figure}[h]
\centering
\includegraphics[clip,width=8.5cm]{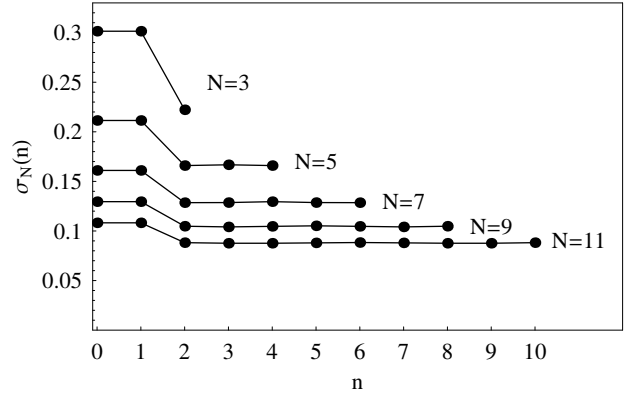}
\caption{Dependence of $\sigma_{N}(n)$ on the position and  system size}
\end{figure}

First we find that $\sigma_{N}(n)$ depends on the position and it has the maximum value at $n=0$ and 1. The dependence of $\sigma_{N}(n)$ on the position 
from 2 to $N-1$ is not clear only from Fig. 1, however 
we find that $\sigma_{N}(n)$$=\sigma_{N}(N+1-n \mbox{\,mod \,} N)$ for any $n$ 
and all values are distinct except a pair $n$ and $(N+1-n) \mbox{\,mod \,} N$.

Next we consider asymptotic behavior of $\sigma_{N}(0)$ for large $N$ to observe the dependence on the system size. Setting $n=0$ in Eq.(\ref{eqn:sigmaF}) gives

\begin{eqnarray}
&& \sigma_{N}^{2}(0) = \frac{1}{N^{3}} 
\left [\sum_{j=0}^{N-1} 
\frac{7\cos \theta_{j}}{3+\cos \theta_{j}}-2 
\right]  \nonumber \\
             && \> + \frac{1}{N^{4}} \left [
\left(
\sum_{j=0}^{N-1}
\frac{1}{3+\cos \theta_{j}}
\right)
\left(
\sum_{j=0}^{N-1}
\frac{15-11\cos \theta_{j}}{3+\cos \theta_{j}}
\right) 
-\sum_{j=1}^{N-1}
\frac{9+4 \cos \theta_{j}+3\cos 2\theta_{j}}{(3+\cos \theta_{j})^{2}}
\right ].  \nonumber \\
\label{eqn:n0}
\end{eqnarray}
On the other hand, we have
\begin{eqnarray}
&& \lim_{N \to \infty} \frac{1}{N}
\sum_{j=0}^{N-1}
\frac{1}{3+\cos \theta_{j}} = \frac{1}{4 \pi}\int_{0}^{4 \pi} \frac{1}{3+\cos x} dx=\frac{1}{2\sqrt{2}}, \\
&& \lim_{N \to \infty} \frac{1}{N}
\sum_{j=0}^{N-1}
\frac{\cos \theta_{j}}{3+\cos \theta_{j}} = \frac{1}{4 \pi} \int_{0}^{4 \pi} \frac{\cos x}{3+\cos x} dx
=1-\frac{3}{2\sqrt{2}}, \\
&& \lim_{N \to \infty} \frac{1}{N}
\sum_{j=0}^{N-1}
\frac{1}{(3+\cos \theta_{j})^{2}} = \frac{1}{4 \pi} \int_{0}^{4 \pi} \frac{1}{(3+\cos x)^{2}} dx
= \frac{3}{16 \sqrt{2}}, \\
&& \lim_{N \to \infty} \frac{1}{N}
\sum_{j=0}^{N-1}
\frac{\cos \theta_{j}}{(3+\cos \theta_{j})^{2}} = \frac{1}{4 \pi} \int_{0}^{4 \pi} \frac{\cos x}{(3+\cos x)^{2}} dx
= -\frac{1}{16 \sqrt{2}}, \\
&& \lim_{N \to \infty} \frac{1}{N}
\sum_{j=0}^{N-1}
\frac{\cos (2 \theta_{j})}{(3+\cos \theta_{j})^{2}} = \frac{1}{4 \pi} \int_{0}^{4 \pi} \frac{\cos 2x}{(3+\cos x)^{2}} dx
= 2 - \frac{45}{16 \sqrt{2}}.
\label{eqn:integrals}
\end{eqnarray}
Using the above results, we obtain asymptotic behavior of the variance $\sigma_{N}^2(0)$ for sufficiently large $N$ as follows: 

\begin{proposition}
For $N \to \infty$, we have
\begin{eqnarray}
\sigma_{N}(0)^2 = \frac{13-8\sqrt{2}}{N^2}+\frac{7 \sqrt{2}-16}{2N^{3}} 
+ o \left( {1 \over N^3} \right).
\label{eqn:approx}	
\end{eqnarray}
\end{proposition}

The above result implies that the fluctuation $\sigma_{N}(0)$ decays in the form $1/N$ as $N$ increases. Fig. 2 shows the comparison between exact values
of $\sigma_{N}(0)$ and the approximate values obtained from Eq. (\ref{eqn:approx}).
We see good agreement between them except $N=3$.

\begin{figure}[h]
\centering
\includegraphics[clip,width=8.5cm]{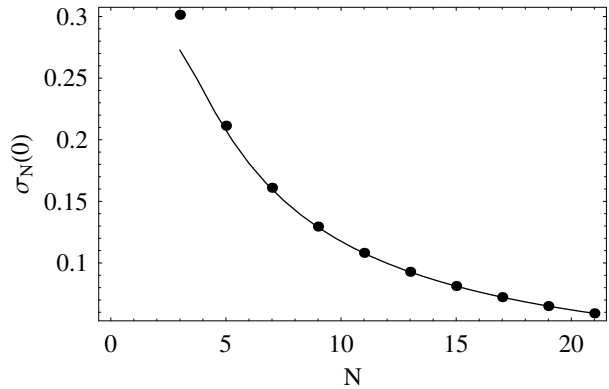}
\caption{Comparison exact values of $\sigma_{N}(0)$ (solid circles) with approximate values (solid line) given by Eq.(46) 
for $N=3,5, \cdots ,21$.}
\end{figure}

\section{Dependence of $\sigma_N(n)$ on Initial States and Parity of System Size}

\subsection{Dependence of $\sigma_{N}(n)$ on Initial States}
\hspace*{1em}
In the previous sections, 
we have shown the results covering only at the case where the system size is odd and the initial state is $|L, 0, 0 \rangle=1$. The time averaged distribution on a circle including odd sites is independent of the initial state.
Is also the temporal standard deviation independent of the initial state?
Since the temporal standard deviation  depends on the position, it is clear that 
the temporal standard deviation depends on the initial state.
We consider here whether the temporal standard deviation depends on the initial state
even if the initial probability distribution is the same.

In order to observe the dependence of the temporal standard deviation on  initial state without
changing the initial probability distribution,
we set $|L,0,0 \rangle$=$\alpha$ and $|R,0,0 \rangle$=${\sqrt{1 - {\alpha }^2}} \> i$ where $\alpha \in [0,1]$. 
As a particular case, we have a symmetrical initial state by setting $\alpha=1/\sqrt{2}$. Thus
we can observe the dependence of $\sigma_{N}(n,\alpha)$ on symmetry break down by changing $\alpha$.
We first present the result for $N=3$.
The standard deviation $\sigma_{3}(n,\alpha)$ for the above initial state
is obtained as follows:
\begin{eqnarray}
\sigma_{3}(0,\alpha) &=& \frac{2\,{\sqrt{46}}}{45},\\
\sigma_{3}(1,\alpha) &=& \frac{2}{45}
{\sqrt{ 96\,{\alpha }^4- 75\,{\alpha }^2 + 25}},\\
\sigma_{3}(2,\alpha) &=& \frac{2}{45}
{\sqrt{ 96\,{\alpha }^4- 117\,{\alpha }^2 + 46}}.
\end{eqnarray}
Fig. 3(a) shows the dependence of $\sigma_{3}(n,\alpha)$ on the parameter $\alpha$.
One clearly finds that $\sigma_{3}(0,\alpha)$ is independent of the initial and
both $\sigma_{3}(1,\alpha)$ and $\sigma_{3}(2,\alpha)$ depend on $\alpha$.
We further find that $\sigma_{3}(1,\alpha)$ is equal to $\sigma_{3}(2,\alpha)$ 
at the symmetrical case $\alpha=1/\sqrt{2}$
and the minimum values are located near $\alpha=1/\sqrt{2}$.

\begin{figure}[htbp]
\centering
\includegraphics[clip,scale=0.5,width=8.5cm]{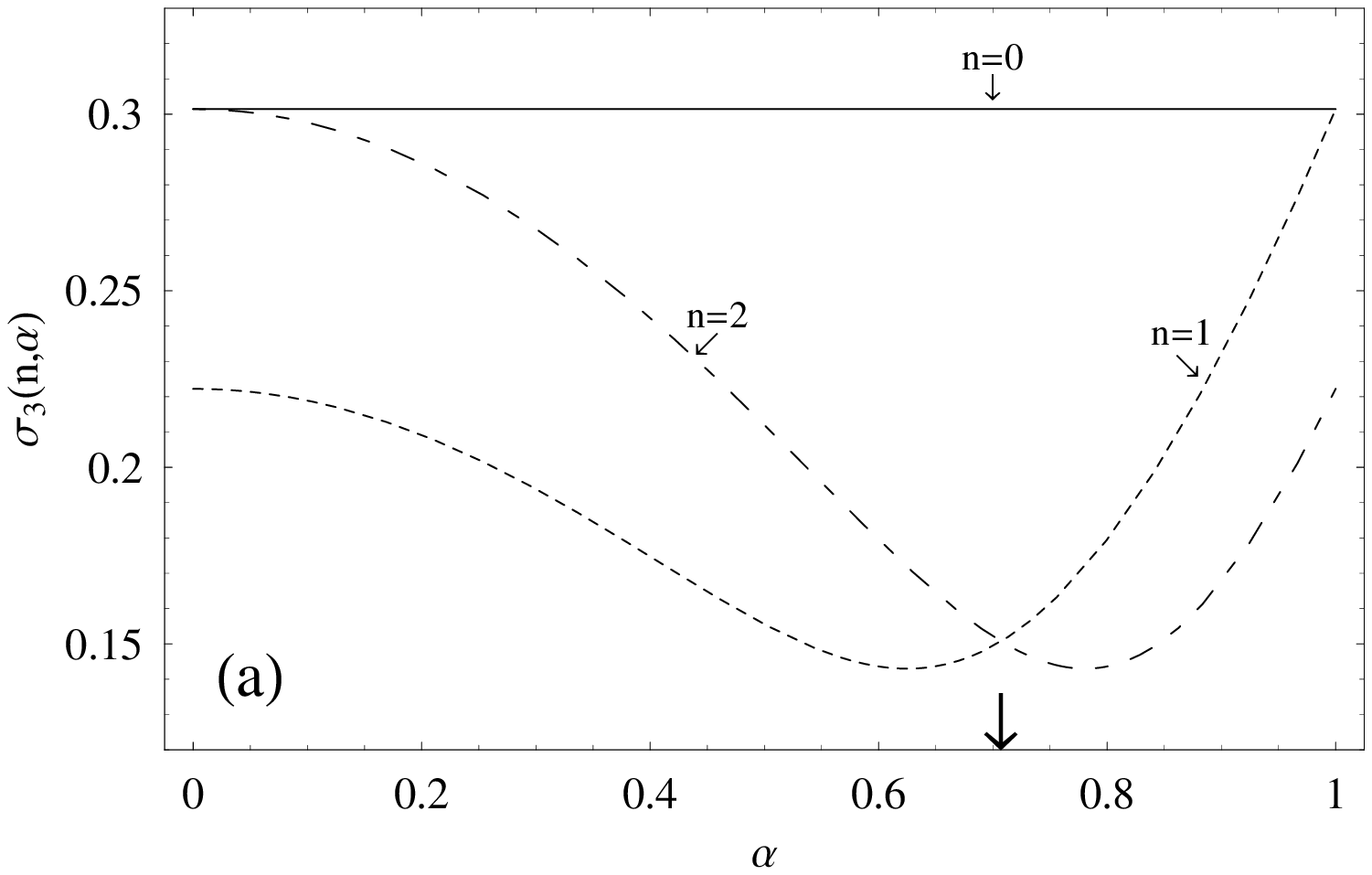}
\includegraphics[clip,scale=0.5,width=8.5cm]{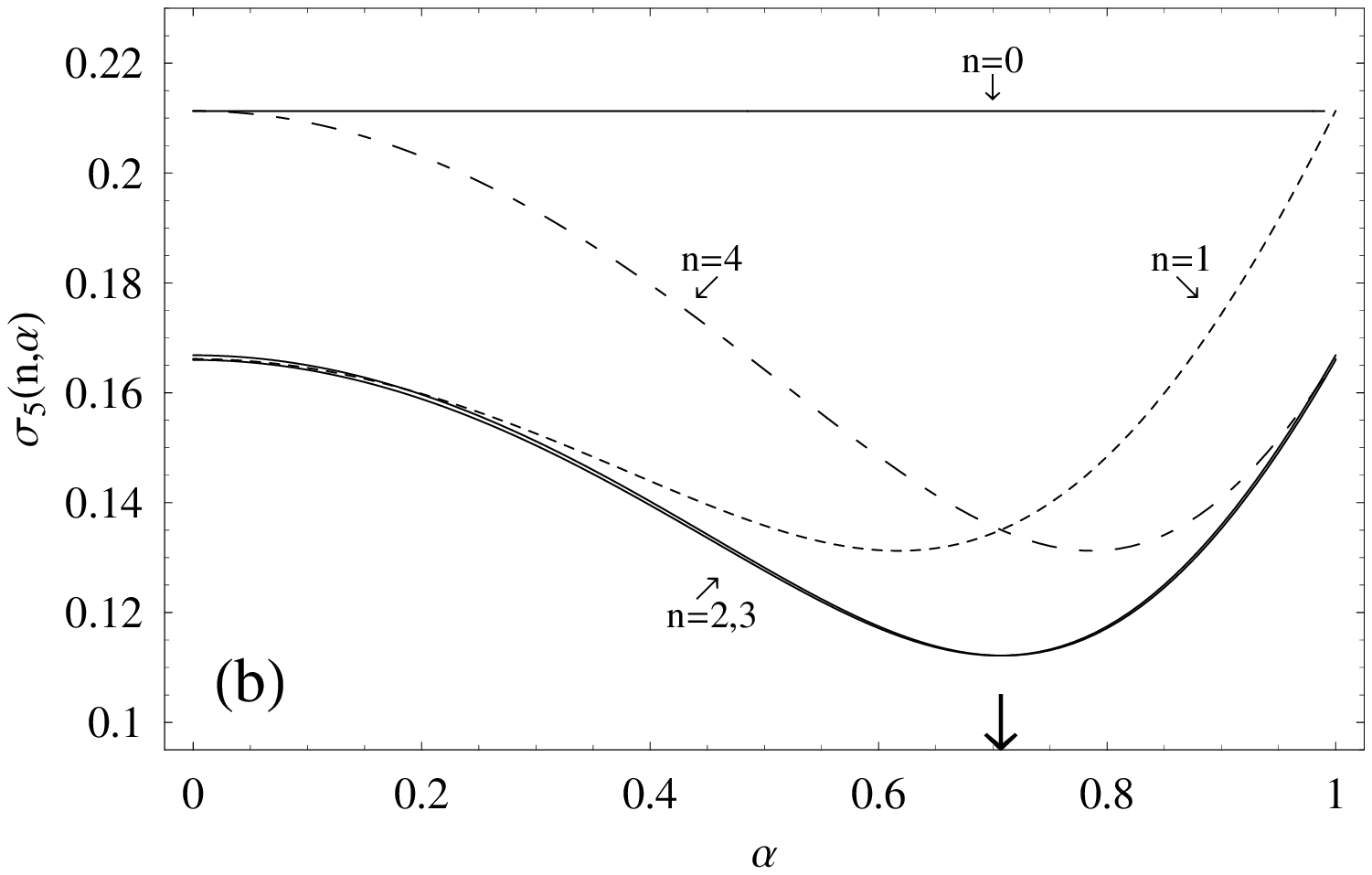}
\includegraphics[clip,scale=0.5,width=8.5cm]{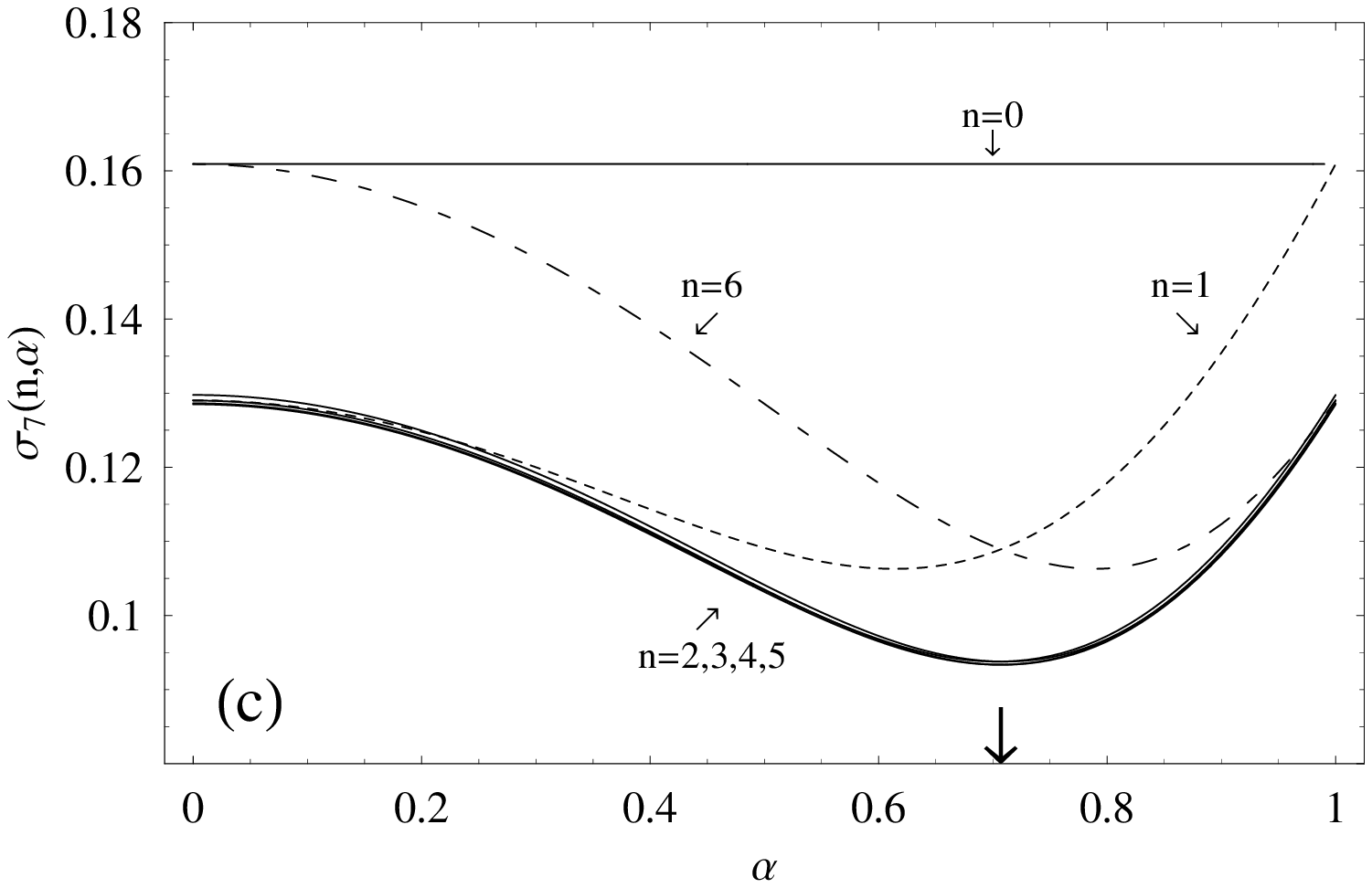}
\caption{Dependence of $\sigma_{N}(n,\alpha)$ on the initial state for (a) $N=3$, (b) $N=5$ and (c) $N=7$.
An arrow indicates $\alpha=1/\sqrt{2}$ near the $x-$axis.}
\label{fig:fig3}
\end{figure}

Fig. 3(b) and Fig. 3(c) show approximate values by setting $T=10^{4}$ in
Eq. (\ref{eqn:average}) and Eq. (\ref{eqn:sigma}) in place of taking a limit for $N=5$ and $N=7$.
Similarly with Fig. 3(a), the temporal standard deviations at the position $n=0$ are independent
of the parameter $\alpha$ and the minimum values are located near $\alpha=1/\sqrt{2}$.
The dependence of $\sigma_{N}(n,\alpha)$ on the position is weak between 2 and $N-2$, and they seems  to be in close on a single line.

\subsection{Temporal Standard Deviation on a Circle with Even Sites}
\hspace*{1em}
We have concentrated our attention into the case where the number of the system size is odd.
The main reason why we avoid even cases is that
the eigenvalues of matrix $M_{N}$ are highly degenerate, and 
the time averaged distribution itself
depends on the position of site and the initial state. 
While the eigenvalues for even cases is exactly obtained by Eq.(\ref{eqn:eigenV}),
there are many possible combinations of eigenvalue which satisfy
$\Delta \theta=0$ (mod 2$\pi$) for even case. Thus we can not calculate $\sigma_{N}(n)$ in the same way with odd cases.
Therefore, we here try to carry out approximate calculations
instead of seeking analytic results.
Fig. 4 shows  the approximate  $\sigma_{N}(n)$ for $N=4,6,\cdots,12$ under 
the same initial condition $|L,0,0 \rangle=1$ and $T=10^4$. 

\begin{figure}[htbp]
\centering
\includegraphics[clip,scale=1,width=8.5cm]{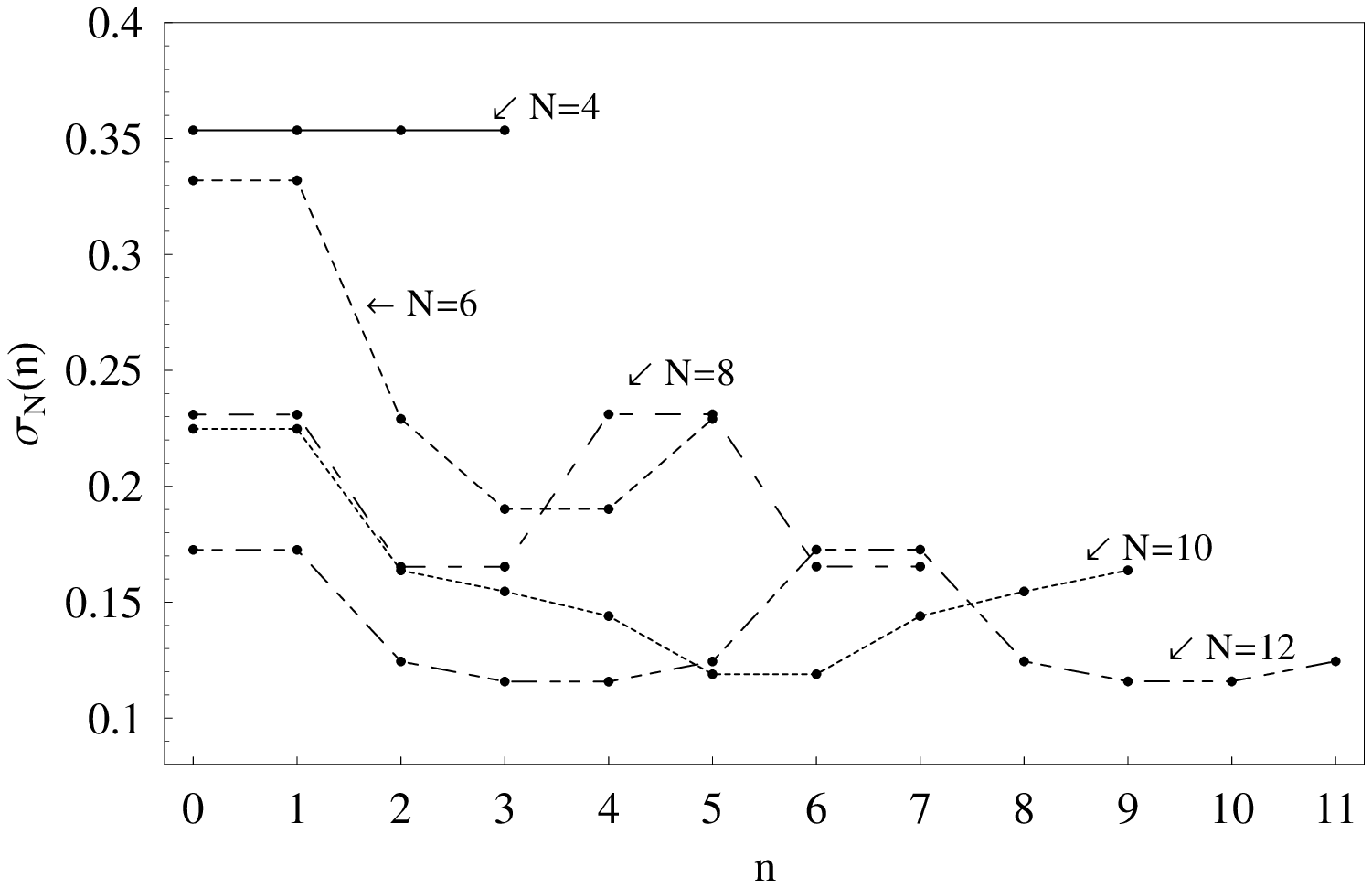}
\caption{Dependence of $\sigma_{N}(n)$ on the position with even  system size.}
\label{fig:sig_even}
\end{figure}

Let us  compare Fig. 1 with Fig. 4. First we find that the values of the temporal standard deviation for fixed $N$
is almost the same except $n=0$, while
the values for even $N$ is strongly dependent on the location of the site in comparison
with odd cases. Second we find that the maximum values of $\sigma_{N}(n)$ for odd $N$ exist only
at $n=0$ and $n=1$ as shown Fig. 1, while  the four same peaks are found for $N=4,8,12$
in Fig. 4. We confirm numerically that this behavior is always observed up to $N=22$ when $N$ is a multiple of 4.

\section{Summary}
\hspace*{1em}
We expressed the temporal standard deviation of the Hadamard walk on a circle with odd sites for a pure initial state $|L,0,0 \rangle =1$ in an analytical form.
The formula is obtained by calculating the production between four Fourier coefficients which come from
production of ``$L$" and ``$R$" states. 
Unlike the time-averaged distribution, the temporal standard deviation depends on the location of the site.
The fluctuations decrease in inverse proportion to the system size for large $N$ and its coefficient was calculated rigorously.

When we set the initial state in the form 
$|L,0,0 \rangle$=$\alpha$ and $|R,0,0 \rangle$=${\sqrt{1 - {\alpha }^2}}\>i$,
the analytical result for $N=3$ and numerical simulations indicated that 
the temporal standard deviation depends on the initial state except $n=0$.
We would speculate that the temporal standard deviation takes the maximum
value at $n=0$ and its value is independent of $\alpha$.
The values of the temporal standard deviation at neighbor site of origin is somewhat larger
than other sites and the remain are almost the same. 
The dependence of the temporal standard deviation on the position for even system size is much complex
in comparing with that for odd system size due to degeneration of eigenvalues. Several peaks are observed when the system size is a multiple of 4.

From the viewpoint of technology, it seen to be of value to consider whether 
the temporal standard deviation is controlled or not. 
It is shown that the time-averaged distribution can not be controlled for odd system size, however, for even system size, it is controlled thanks to the degenerate of eigenvalues. Our results show that the standard deviation can be controlled.
First the dependence of temporal standard deviation on location is used. Second the dependence on
initial state is also used. However we note here that $\sigma_{N}(0)$ for $N=3$ is independent
of the initial state. In this meaning, its control is limited.
\par
\
\par\noindent
{\bf Acknowledgments}

This work is partially financed by the Grant-in-Aid for Scientific Research (B) (No.12440024) of Japan Society of the Promotion of Science.


\begin{thebibliography}{0}

\bibitem{1} J. Kempe, ``Quantum random walks - an introductory overview'', {\it  Contemporary Physics}, {\bf 44}, 307 (2003).

\bibitem{2} A. Ambainis, E. Bach, A. Nayak, A. Vishwanath and J. Watrous, ``One-dimensional quantum walks'', in Proceedings of the 33rd Annual ACM Symposium on Theory of Computing, 37 (2001).

\bibitem{3} N. Konno, ``Quantum random walks in one dimension'', {\it Quantum Information Processing}, {\bf 1}, 345 (2002).

\bibitem{4} D. Aharonov, A. Ambainis, J. Kempe, and U. V. Vazirani, ``Quantum walks on graphs'', in Proceedings of the 33rd Annual ACM Symposium on Theory of Computing, 50 (2001).

\bibitem{5} M. Bednarska, A. Grudka, P. Kurzy\'nski, T. Luczak and A. W\'ojcik, ``Quantum walks on cycles'', quant-ph/0304113.


\bibitem{6} R. B. Schinazi, {\it Classical and Spatial Stochastic Processes} 
(Birkh\"auser, Boston, 1999).


\end{thebibliography}
\end{document}